\documentclass{emulateapj}
\usepackage{graphicx}
\usepackage{threeparttable}
\usepackage{hyperref}
\usepackage{color}

\newcommand{\xmm}{{\it XMM-Newton} }

\newcommand{\rosat}{{\it ROSAT} }

\newcommand{\swift}{{\it Swift} }
 
\renewcommand{\arcsec}{\mbox{$^{\prime\prime}$} }

\newcommand{\logxi}{erg cm s$^{-1}$}
\newcommand{\flux}{{erg~cm$^{-2}$~s$^{-1}$ }}

\newcommand{\nh}{cm$^{-2}$}

\newcommand{\ergs}{${\rm erg \ cm^{-2} \ s^{-1}}$ }
\newcommand{\erg}{${\rm erg \ s^{-1}}$ }

\def\ts     {\thinspace}

\def\kms    {\ifmmode{{\rm \ts km\ts s}^{-1}}\else{\ts km\ts s$^{-1}$}\fi}
\def\msol   {\ifmmode{{\rm M}_{\odot}}\else{M$_{\odot}$}\fi}
\def\lsol   {\ifmmode{{\rm L}_{\odot}}\else{L$_{\odot}$}\fi}
\def\zsol   {\ifmmode{{\rm Z}_{\odot}}\else{Z$_{\odot}$}\fi}

\def\ltsima{$\; \buildrel < \over \sim \;$}
\def\simlt{\lower.5ex\hbox{\ltsima}}
\def\gtsima{$\; \buildrel > \over \sim \;$}
\def\simgt{\lower.5ex\hbox{\gtsima}}

\newcommand{\msun}{{\rm\,M$_\odot$}}

\newcommand{\src}{{\rm\,GSN 069~}}
\newcommand{\srcs}{{\rm\,GSN 069}}

\begin{document}

\title{
A long decay of X-ray flux and spectral evolution in the supersoft Active galactic nucleus GSN 069}
\author{X. W.~Shu\altaffilmark{1}, 
S. S.~Wang\altaffilmark{1}, 
L. M.~Dou\altaffilmark{2}, 
N. Jiang\altaffilmark{3},
J. X.~Wang\altaffilmark{3}, 
T. G.~Wang\altaffilmark{3},
}
\altaffiltext{1}{Department of Physics, Anhui Normal University, Wuhu, Anhui, 241000, China; 
xwshu@mail.ahnu.edu.cn}
\altaffiltext{2}{Center for Astrophysics, Guangzhou University, Guangzhou 510006, China}
\altaffiltext{3}{CAS Key Laboratory for Researches in Galaxies and Cosmology, Department of Astronomy, University of Science and Technology of China, Hefei, Anhui 230026, China}



\begin{abstract}

GSN 069 is an optically identified very low-mass AGN which
shows supersoft X-ray emission. 
The source is known to exhibit huge X-ray outburst, with flux increased 
by more than a factor of $\sim240$ compared to the quiescence state. 
We report its long-term evolution in the X-ray flux and spectral variations over a time-scale of $\sim$decade, 
using {both} new and archival X-ray observations from the \xmm and \swift. 
The new \swift observations detected the source in its lowest level of X-ray activity since outburst, 
a factor of $\sim$4 lower in the 0.2-2 keV flux than that obtained with the \xmm observations 
nearly 8 years ago. 
Combining with the historical X-ray measurements, we find that the X-ray flux is 
decreasing slowly. 
There seemed to be 
spectral softening associated with the drop of X-ray flux. In addition, we find evidence for the presence of 
{a} weak, variable hard X-ray component, in addition to the dominant thermal blackbody emission
reported before. 
The long decay of X-ray flux and spectral evolution, as well as the supersoft X-ray spectra, 
suggest that the source could be a tidal disruption event, though a highly variable AGN cannot be fully 
ruled out. Further continued X-ray monitoring would be required to test the 
TDE interpretation, through better determining the flux evolution in the decay phase.
\end{abstract}


\section{Introduction}

The typical  X-ray  spectrum  of  Seyfert  1  galaxies consists of 
a hard power-law emission with an exponential cutoff at an energy above $\sim$100 keV 
(Nandra \& Pounds 1994), plus a soft excess below $\sim$2 keV.  
It is believed that the hard powerlaw component is produced by 
the inverse Compton scattering of the seed photons from a cold accretion disk 
in a hot, optically thin corona (Sunyaev \& Titarchuk 1980). 
The nature of soft excess emission is quite elusive to date. 
Although it usually can be fitted with a blackbody, the temperature appears to 
be roughly constant ($kT=0.1\sim0.2$ keV), 
and is much too high to be explained by the standard 
accretion disk model (Gierli{\'n}ski \& Done 2004; Ai et al. 2011). 

Recently, Terashima et al. (2012) reported the discovery of a {remarkable} supersoft AGN 
(J1231+1106), which completely lacks emission at energies $\simgt$2keV.   
Its X-ray spectrum can be represented purely by a soft thermal component with a
blackbody temperature of $kT\sim$ 0.13--0.15 keV, by analog with the accretion disk 
dominated spectrum typically seen in the high/soft state of X-ray binaries (XRBs). 
Note that similar supersoft emission was also reported in the AGN GSN 069 (Miniutti et al. 2013, M13), 
and RX J1302+2746 (Sun, Shu \& Wang 2013, Sun13), but with a lower blackbody temperature 
roughly consistent with the disk model prediction. 
{While `supersoft' AGNs with the soft X-ray photon index $\Gamma_{\rm soft}\simgt4$ have 
been initially identified with the ROSAT data 
(e.g., Boller et al. 2010), 
they are not as extreme in the X-rays as the AGNs we mentioned according to the apparent lack of 
hard X-rays above $\sim$2 keV.}
The discovery of this {special} class of AGNs is {thus} important to illuminate accretion physics 
in a poorly explored regime of parameter space (Ho et al. 2012; Shu et al. 2017).

While {such} supersoft X-ray spectra are rare among AGNs, they are more commonly seen in tidal disruption event (TDE, 
see Komossa 2015 for a review). 
Such an event occurs when a star approaches too close to a SMBH and is tidally disrupted (Rees 1988).  
Subsequently fall-back accretion can form a transient accretion disk whose 
thermal emission {falls into} the X-ray band for a BH mass $M_{\rm BH}\simlt10^6$\msun, 
resulting in a supersoft X-ray spectrum.  
Indeed, two out of three known supersoft AGNs, 
J1231+1106 and GSN 069, display large amplitude X-ray variability by a factor of $>$15 to 240, respectively. 
This leads to {a} recent work to claim that J1231+1106 could be associated with a TDE, though its flux decay seems  
much slower than most other TDE candidates (Lin et al. 2017a). 
The nature of soft X-ray {burst} in GSN 069 is still poorly constrained. 
While observed large variability of more than a factor of 240, further one-year (2010-2011) 
Swift monitoring campaign did not find significant decay in the flux, which is inconsistent with typical TDEs. 

In this Letter, we report the results from  
archival Swift and XMM observations of \src since 2011, which have not been analyzed before. 
In particular, we have obtained new Swift follow-up observations of the source at the end of 
2017, revealing a significant decrease in the X-ray flux. 
Table 1 lists the details of all X-ray observations. 
In Section 2, we describe the observations and data reductions. In Section 3, 
we present detailed analysis and main results. Discussions on the source properties 
are given in Section 4. 

 \begin{figure*}[t]
\centering
 \includegraphics[scale=0.65,angle=0]{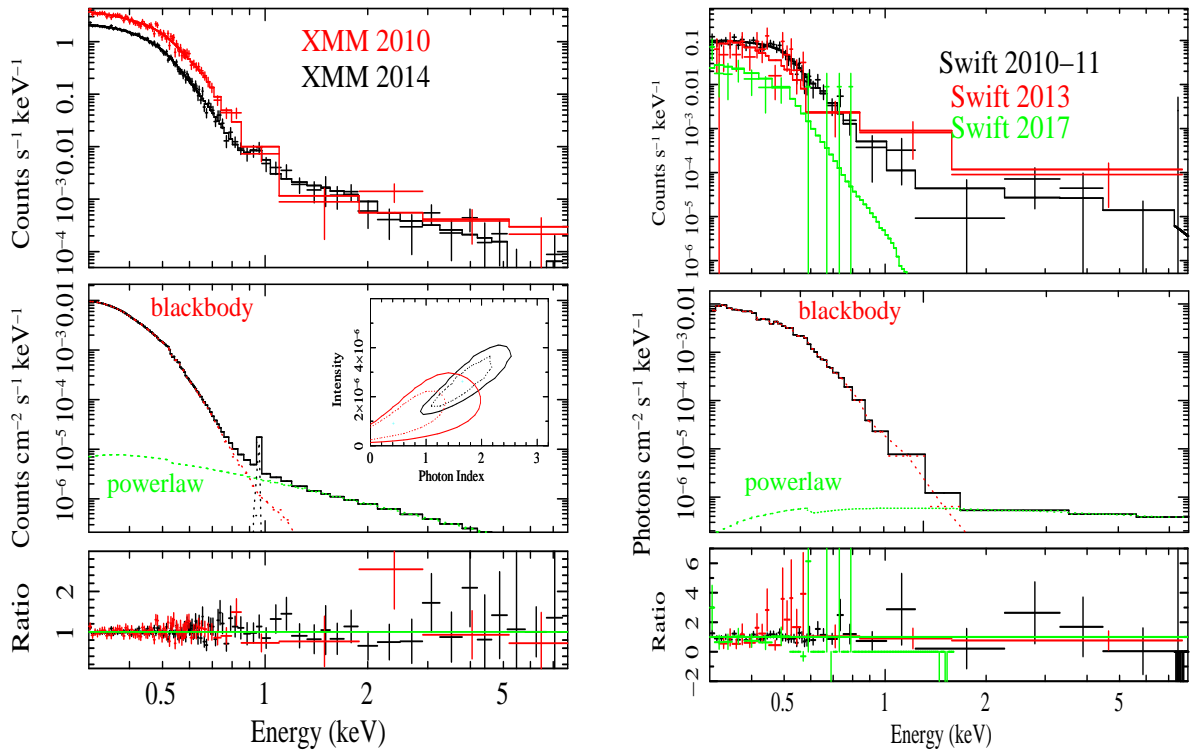}
 \caption{
{\it Left:} \xmm PN spectra for \src {along with the best-fitted models} (top panel). 
The corresponding data/model ratios are shown in bottom panel. 
{The inset panel shows the joint 68\% and 90\% confidence contours of the 
photon index vs. normalized flux for the hard powerlaw component.
{\it Right:} The same as left but for the \swift data. 
We present in each panel the detailed spectral models for 
the {\it XMM}2014 (left) and \swift2011 (right) data.} 
}
\label{fig:youngsed}
\end{figure*}
\section{Data analysis}
\label{sec:data}

\begin{table} \caption{
X-ray observations of \src
\label{tab:bestfit_neut}
}
\begin{center}
\begin{tabular}{llllccc}
\hline
Instrument & Date  &   Flux$^{\dag}$ & Count rates & Refs. \\
           &    &   & ($10^{-2}$ cts/s) & \\
\hline
RASS & 1990 & $<0.05$ &  $\dots$ &  (1)  \\
ROSAT-PSPC & 1993-07-13  & $<0.009$  & $\dots$& (1) \\
ROSAT-PSPC & 1994-06-29 & $<0.007$ & $\dots$ & (1)\\
XMM-PN & 2010-07-14    & 2.4$\pm$0.7  & $\dots$ & (1)\\
SWIFT-XRT & 2010-08-26 & 2.1$\pm$0.6  & 3.73$\pm$1.14 & (1)\\
SWIFT-XRT & 2010-08-27 & 2.1$\pm$0.3  & 3.42$\pm$0.41 & (1)\\
SWIFT-XRT & 2010-10-27 & 1.8$\pm$0.3  & 3.12$\pm$0.33 & (1)\\
SWIFT-XRT & 2010-11-24 & 1.6$\pm$0.3   & 2.78$\pm$0.31 & (1)\\
XMM-PN   & 2010-12-02 &  2.03$\pm$0.03 & $\dots$   & (1)\\
SWIFT-XRT & 2010-12-22 & 1.4$\pm$0.3 & 2.30$\pm$0.27 &(1)\\
SWIFT-XRT & 2011-01-19 & 1.4$\pm$0.3 & 2.12$\pm$0.27 &(1)\\
SWIFT-XRT & 2011-02-16 & 1.5$\pm$0.3 & 2.51$\pm$0.30 &(1)\\
SWIFT-XRT & 2011-04-25 & 1.4$\pm$0.3 &2.56$\pm$0.25 &(1)\\
SWIFT-XRT &2011-05-23 & 1.7$\pm$0.3 & 3.09$\pm$0.27 & (1)\\
SWIFT-XRT & 2011-06-20 & 1.8$\pm$0.3 & 3.03$\pm$0.28 &(1)\\
SWIFT-XRT & 2011-07-17 & 1.4$\pm$0.3 & 2.41$\pm$0.22 & (1)\\
SWIFT-XRT & 2011-08-15 & 2.0$\pm$0.3 &  3.59$\pm$0.36 &(1)\\ 
SWIFT-XRT & 2011-08-18 & 1.9$\pm$0.3 & 3.26$\pm$0.62 &(1)\\
 SWIFT-XRT & 2013-09-22 & 0.84$\pm$0.3      &  1.74$\pm$0.24 & (2) \\
 SWIFT-XRT & 2013-09-29 & 0.74$\pm$0.3 & 1.28$\pm$0.22 & (2) \\
 XMM-PN & 2014-12-05 & 1.3$\pm$0.01 & $\dots$  &(2) \\
 SWIFT-XRT & 2017-12-09 & 0.57$\pm$0.3 & 0.61$\pm$0.20 & (2) \\
 SWIFT-XRT & 2017-12-19 & 0.69$\pm$0.3 & 0.98$\pm$0.26 & (2) \\
 SWIFT-XRT & 2017-12-25 & 0.61$\pm$0.3 & 0.67$\pm$0.19 & (2) \\
 \hline
\end{tabular}
\end{center}
{\bf Note--} $^{\dag}$ The 0.2--2 keV flux in units of $10^{-12}$\ergs. 
References: (1) M13; (2) this work. 
\end{table}

\subsection{XMM observations}
\xmm pointed observations of \src were performed twice on 2010 and 2014 
(hereafter {\it XMM} 2010 and {\it XMM} 2014). 
While the {\it XMM} 2010 data have been presented by M13, here we re-analyzed 
the spectra in detail in the context of the long-term evolution of 
the X-ray flux.  
The \xmm data were reprocessed with the
Science Analysis Software version 16.0.0, using the calibration
files as of 2016 December. 
The epochs of high background events were examined
and excluded using the light curves in the energy band above
12 keV. 
We used principally the PN data,
which have much higher sensitivity, using the MOS data only to
check for consistency when needed. We extracted the source spectra using a circular 
region with a radius of 40\arcsec centered at the source position.
Background spectra were made from source-free areas on the
same chip using four circular regions identical to the source
region. 
We {grouped} the spectra to have at least 20 counts in each bin so as
to adopt the $\chi^2$ statistic for the spectral fits. 
The following spectral fittings are performed in the 0.3-8 keV range 
using XSPEC version 12. All statistical errors given
hereafter correspond to 90\% confidence for one interesting
parameter ($\Delta\chi^2=2.076$), unless stated otherwise.

\subsection{Swift observations}

Totally 16 archival \swift observations of \src are available, among which 
13 are made between 2010 March and 2011 July, and 2 on 2013 September. 
Some results from the \swift 2010-2011 observations have been presented in M13, while the 2013 data 
have not been reported yet. 
In addition, at our request, we have obtained 3 \swift follow-up observations on 2017 December. 
All \swift observations were reduced with FTOOLS 6.19 and updated calibration
files (released on 2017 November 13). The X-ray telescope
(XRT; Burrows et al. 2005) was operated in Photon Counting
mode. We reprocessed the event files with the task
{\tt xrtpipeline} (version 0.13.2). The source was weakly 
detected in all observations, with a count rate in the range 
$\sim0.6-3.7\times10^{-2}$ cts/s {in 0.2--2 keV} (Table 1). Source spectrum was extracted using a source region of radius
20\arcsec. The background was estimated in an annulus region centered on the source position,  
with an inner radius of 30\arcsec and outer radius of 50\arcsec. 
Due to the low counts, we {grouped} the data to have at least 1--5 counts in each bin 
and adopt mainly the $C$-statistic for the \swift spectral fittings.



\begin{table*} \caption{
Spectral fitting results for historical \xmm and \swift data.
\label{tab:bestfit_neut}
}
\begin{center}
\begin{tabular}{llllccc}
\hline
 Model component & Parameter  &  \xmm & \xmm & \swift & \swift   &   \swift\\
            & &  (2010) & (2014) & (2010-2011) & (2013) & (2017)\\ 
 \hline
&&   \\
Blackbody & $kT_{\rm BB}$ (eV)  &    59$\pm$3 &   49$\pm$1              &  54$\pm$5  & 45$^{+24}_{-17}$    & 43$^{+25}_{-13}$\\
Power-law & $\Gamma$      &  $0.43^{+1.03}_{-1.36}$ & $1.67_{-0.55}^{+0.66}$  & 0.43(fixed) & 1.67(fixed) & $\dots$\\
Neutral absorber  & N$_\mathrm{H}$~($\times10^{20}$\nh)& $<3.2$  & $1.4^{+0.9}_{-0.8}$  & $4.0^{+3.9}_{-3.4}$ & $<$44.2  & $<$40\\
Ionised absorber &  N$_\mathrm{H}$~($\times10^{22}$\nh)&   2.3$^{+0.7}_{-1.0}$  & 3.0$^{+0.02}_{-1.2}$ & $\dots$ & $\dots$ &   $\dots$\\
                 & log$\xi$~(\logxi) & 0.3$^{+0.03}_{-0.08}$   & 0.08$^{+0.22}_{-0.38}$ & $\dots$ & $\dots$  & $\dots$  \\    
\\

F$_{(0.2-2)\mathrm {keV}}$ & Absorbed~($10^{-12}$\flux) & $1.95\pm0.03$  & $1.3\pm0.01$ & $1.8\pm0.2$ & $0.74\pm0.3$ & $0.63^{+0.2}_{-0.09}$\\
F$_{(2-10)\mathrm {keV}}$  & Absorbed~($10^{-14}$\flux)  & $3.18^{+1.2}_{-2.5}$  & $0.71^{+0.27}_{-0.35}$ & $2.42^{+2.2}_{-1.7}$ & $4.9^{+2.4}_{-2.7}$  & $\dots$\\
Statistics  & $\chi^2/dof$ &   70.9/65   & 98.1/93 & 25.6/37 & 20.4/14$^{\rm a}$ & 43.6/57$^{\rm a}$ \\
 \hline
\end{tabular}
\end{center}
{\bf Note--}
$^{\rm a}$ $C$-statistic is used due to low counts.  
\end{table*}

\section{Result}

 \subsection{X-ray spectral analysis}
 Despite the factor of $>240$ increase in flux between the \rosat observation and 
 the \xmm observation that first discovered the source in the X-rays on 2010, 
 a 1 year monitoring campaign of \src with \swift beginning in 2010 August revealed 
 no significant flux variations by a factor of $\simlt1.5$ (M13).  
 Here, we present a detailed and uniform analysis of all the archival X-ray observations, 
 especially for those that are not given in M13.  
 In addition, we present the new results from the most recent \swift follow-up observations,  
 with {the} aim to further investigate the evolution in the X-ray flux and/or the spectral changes 
 on longer time-scale.

   Considering the very soft spectra known for the source, 
   as a first step, we fitted a pure thermal blackbody model ({\sc zbbody} in XSPEC) to the {\it XMM}2014 spectrum, 
   which has the best spectral quality thanks to its long exposure.  
   The model gives very poor description to the spectrum, with a $\chi^2$/dof=228.3/101. 
   The spectral ratio shows that there 
   is a considerable excess of flux at the high energies above 2 keV. 
   The net count rate in the 2--8 keV band is $8.4\pm2.8\times10^{-4}$ cts/s, 
   corresponding to $\sim$54$\pm$20 background subtracted counts. 
   Such an excess is likely due to the {underlying} hard power-law component, which is commonly seen 
   in the X-ray spectrum of nearby Seyfert 1 galaxies (e.g., Bianchi et al. 2009).  
   {The fit} by adding a power-law model is much better, with $\chi^2$/dof=149.4/99 
   and a best-fit photon index $\Gamma=1.98$. 
   Note that there is still curvature in the spectrum between $\sim$0.6 and 1 keV, 
   which can be attributed to the warm absorbing gas along the line of sight. 
   We therefore included a warm absorber model {\sc zxipcf} (Reeves et al. 2008), and 
   obtained a significantly improved fit with $\chi^2$/dof=104.3/95, i.e., 
   a decrease in $\chi^2$ of 45.1 for four more extra free parameters. 
   The warm absorber has a column density 
   $N_H=3.0^{+0.02}_{-1.2}\times10^{22}$ cm$^{-2}$ and ionization parameter log$\xi=0.08^{+0.22}_{-0.38}$ \logxi.  
   The photon index now becomes flatter, with $\Gamma=1.67^{+0.66}_{-0.55}$. 
   Our final model also includes a Gaussian line to take into account the excess emission 
   at $\sim1$ keV. 
   { 
   Since the line is unresolved, we fixed its width at a value of 5 eV (much less than the 
   PN spectral resolution), and obtained a best-fit energy of $E=0.97\pm0.04$ keV.  
    Although the residuals vanish, the overall value of $\chi^2$ is reduced by only 5.9 for the
   addition of two free parameters, suggesting that the feature is not statistically significant.  
   If it were real, the line would be due to  Ne{\sc X} Ly$\alpha$ emission line which has a rest-frame 
   energy of 1.022 keV, which can be verified with other data.   
   }

   In order to obtain constrains on the spectral evolution for \srcs, we also performed 
   such an analysis on the \xmm spectrum taken on 2010. 
   We used the same baseline model as that used to fit the {\it XMM}2014 data. 
   A comparison of the X-ray spectrum between the two \xmm observations is shown 
   in Figure 1(left), and the spectral fitting results are shown in Table 2. 
   Within the statistical errors, the parameters for the ionised absorber are consistent 
   among the two data sets. 
   {The power-law slope 
   appears to be flatter in comparison with the {\it XMM}2014 data, 
   but still marginally consistent with each other within 90\% uncertainty (see the 
   inset panel in Figure 1) due to poor statistics. 
   }
   In addition, 
   the thermal temperature for the blackbody component is higher. 
   Note that the fitting results for the {\it XMM}2010 
   spectrum are formally consistent with that reported in M13, except for the 
   hard X-ray emission {for} which only upper limits could be obtained. 

   We now performed similar analysis on the historical 2010 and 2013 \swift spectra, 
   as well as the 2017 \swift spectra we obtained. 
   Unfortunately, the spectral signal-to-noise (S/N) ratios for most of the individual \swift spectra 
   are not sufficient to perform meaningful fits. Thus, in order to obtain a spectrum with better 
   S/N ratio, we combined the spectra from individual observations using the FTOOLS task {\tt addascaspec}. 
   Since the source flux is likely changed during the epoch 2010-2017, we added spectral files separately 
   from 
   the 13 \swift observations {obtained} from a 1 year monitoring campaign {between} 2010-2011 (\swift 2011), 
   2 observations taken on 2013 (\swift 2013), and 3 observations taken on 2017 (\swift 2017).  
   We obtained 820$\pm$30, 48$\pm$7 and 23$\pm$5 net counts in the 0.3-8 keV for the \swift 2011, 2013 
   {and} 2017, respectively. 
   The spectra appear very soft, with most of the counts below $\sim$0.7 keV. 
   For the \swift 2011 and 2013 observations, we also detected $5\pm3$ and $2\pm1$ net counts at 
   energies above 1.5 keV, but they are too low to be used to constrain the hard X-ray power-law index effectively. 
   We therefore fixed the corresponding parameters to that measured from the closest \xmm observations. 
   As might be expected, the ionised absorber could not be well constrained due to {the poor} S/N ratio 
   of the \swift data, so it was not included in the fits. 
   As for the \xmm observations, we present in Figure 1 (right) a comparison of the 
   \swift spectrum between the three epochs. 
   All of the spectral fitting results for the \swift data are shown in Table 2.


\begin{figure}[tp]
\centering
\includegraphics[scale=0.65,angle=0]{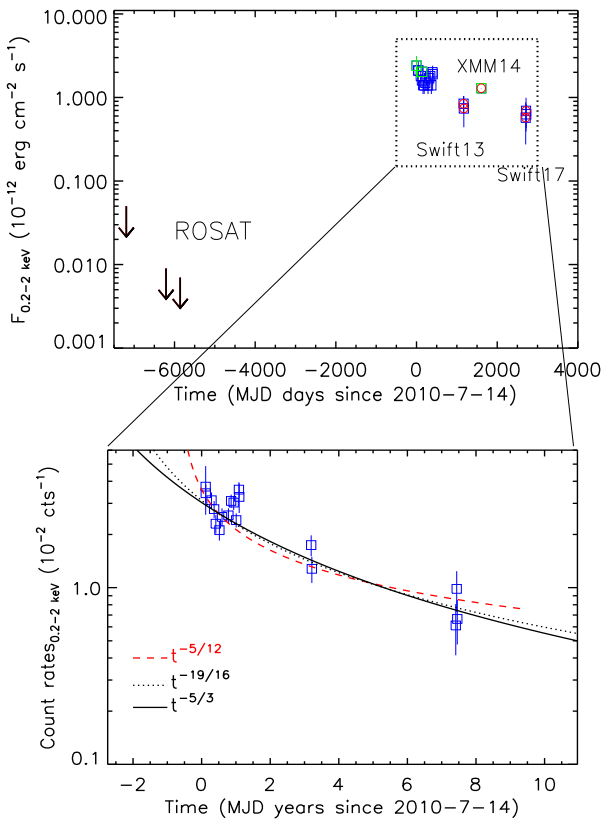}
 \caption
{
{\it Upper panel:} Historical flux lightcurve in the 0.2-2 keV for \srcs. 
The \swift data are shown in blue squares, while {\it XMM} data are shown 
in green. The new data (after 2013) that have not been presented in M13 are highlighted 
in red circles. We also plotted the upper limits (downward arrows) from ROSAT observations.  
{\it Lower panel:} The evolution of the \swift count rates in the 0.2--2 keV. 
{The best-fitting model with a $(t-t_{D})^{-5/12}$, $(t-t_{D})^{-19/16}$, and $(t-t_{D})^{-5/3}$ 
decay law is shown in red dashed, black dotted and solid line, where $t_{D}\simeq0.68, 3.29$ and 5.62 
in units of years before 2010-07-14, respectively.}
}
\end{figure}


     
 \subsection{The long-term flux variability}

 {Figure 2 (upper panel) plots the long-term evolution of the observed X-ray flux of \src in the 0.2-2 keV.  
 Since we are investigating the variability behaviour on longer time-scale,  
 the upper limits from the earlier \rosat observations are also included, 
 building a light curve with a time span of more than two decades}. 
 It is evident that the source flux appears to decay slowly with time since the outburst. 
 The 0.2-2 keV X-ray flux had decreased to {a} historical low state level in the recent \swift2017 observation, 
 by a factor of $\sim$4 lower than the brightest flux found in the {\it XMM}2010 observation. 
 While a weak re-brightening during the decay period is recorded in the {\it XMM}2014
 data, the flux in fact {has dropped by a factor of $1.5$ in comparison with the {\it XMM}2010 observation}
 .  

In order to study the variability behavior in a relatively model independent
way, we replaced the flux with count rates but for all historical \swift observations only. 
{The lower panel of Figure 2} shows the long-term evolution of the \swift count rates {in the 0.2--2 keV}, 
which displays a similar slow decay, by a factor of a few over $\sim$decade in time. 
Such an evolution of the X-ray flux depends on many factors, and cannot trivially be explained 
without knowing the details of the accretion process. 
We attempted to explore the fiducial model assuming a power-law decline, i.e., 
 $\sim(t-t_{\rm D})^{-\alpha}$, which has been commonly used to fit the lightcurve of TDEs. 
{ 
It should be noted that given the low effective area of {\swift} below 0.3 keV and extremely soft X-ray spectrum 
for the source, the true uncertainty in the light curve fit is larger than that implied by using count rate as a proxy 
for the flux, which may be affected by the choice of energy band, absorbing column density and intrinsic spectral 
shape of the source (e.g., blackbody temperature) and its variability. 
}

{While the $t^{-5/3}$ decay law represents the typical evolution trend for many thermal TDEs,
it has been suggested that the actual luminosity evolution may depend on many factors
such as the structure of the disrupted star, and the fraction of fallback materials is directly
translated into an accretion luminosity (e.g., Lodato et al. 2015).
When the luminosity is dominated by the accretion of the fallback material,
Lodato \& Rossi (2011) have shown that at late times the optical/UV lightcurves are proportional to
$t^{-5/12}$, and thus substantially flatter than the $t^{-5/3}$ behaviour. 
Strubbe \& Quataert (2009) showed that the emission at early times can contain some contribution from 
a radiatively driven wind or outflow, leading to a $t^{-5/9}$ decline.  
In addition, by assuming a viscous disk accretion (viscous timescale is much longer than the fallback
one),
Cannizzo et al. (1990) calculated that the evolution of luminosity follows
a powerlaw with index of $\alpha=19/16$.
Based on simple powerlaw model fits, Auchettl et al. (2017) found that the evolution of X-ray emission 
for a large sample of TDEs appears to follow either the commonly used $t^{-5/3}$ accretion, 
or the more shallower decline law of $t^{-5/12}$.  


It is important to note that the current data cannot tightly constrain
the value of the decay index, since the actual TDE time, $t_D$, is largely unknown.
Therefore we fitted the count rate lightcurve by assuming various $t_D$ before the initial outburst, 
namely, between 2010-07-14 ($t_{\rm max}$) where the highest X-ray flux was observed for the source and 
1994-06-29 when the last ROSAT observation was performed. 
We created a grid of $t_D$ in time bin of 50 days between the above time interval, and
performed the powerlaw fits at each $t_D$, while allowing the normalisation and the
powerlaw index $\alpha$ to be free parameters. 
In the fits, we also set constrain on the peak X-ray luminosity to be less than $10^{44}$ erg/s, 
which has been observed for most TDEs discovered so far (e.g., Komossa et al. 2015). 
{This is approximately the Eddington luminosity for GSN 069 for a BH mass of $M_{\rm BH}\sim10^6$\msun 
(M13).}

Figure 3 (a) and (b) shows the $\chi^2$ values and powerlaw indices from best-fits as 
a function of $t_D$ (in units of years referred to 2010-07-14, $t_{\rm max}$), respectively. 
As one would expect, the $t_D$ and powerlaw decay index are poorly constrained, and
no significant difference in the $\chi^2$ statistics is found in a wide $t_D$ range
within $\sim$1--10 years before $t_{\rm max}$. Specifically,
the difference in the $\chi^2$ is only 2.58 (at a $72.47$\% level) between the canonical $t^{-5/3}$
and the more shallower $t^{-5/12}$ decline. 
The corresponding TDE time $t_D$ for both cases is, however, very different, which 
is $\sim$0.68 year and 5.6 years before
the peak of the luminosity seen by {\it XMM}, respectively. 
Although current data can not constrain models well because of 
the degeneracy between the $t_D$ and powerlaw index, 
it is likely that the source will continue to decline in flux (Figure 2, lower panel). 
Thus future sensitive X-ray observations with higher S/N are required to track its flux 
and provide further observational constraints on the decay curve. 

}


\begin{figure}[tbp]
\centering
\includegraphics[scale=0.45,angle=0]{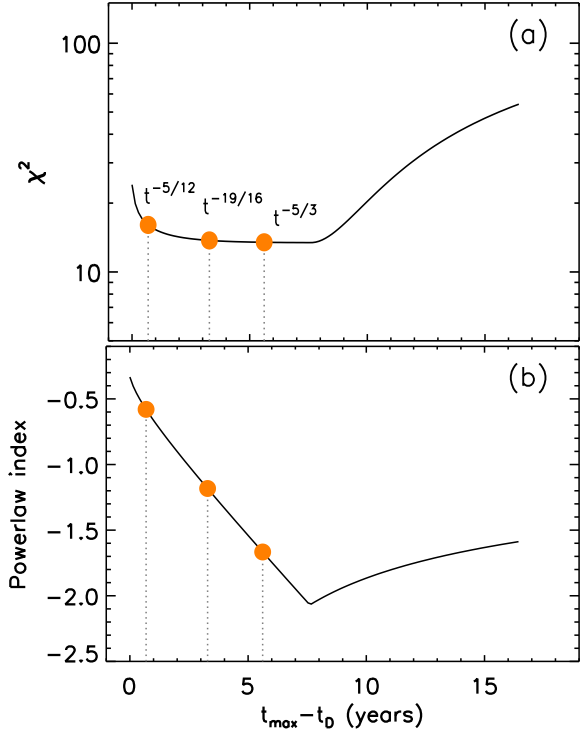}%
\caption{
{\it Upper panel}:
The best-fitting $\chi^2$ values are plotted against the actual time of TDE from 2010-07-14 to
1994-06-29 (or $\sim$0-16 years before 2010-07-14).
The yellow solid circles represent the best-fit value of $t_D$ at which
the decay index is $\alpha=-5/12$, -19/16 and -5/3, respectively.
{\it Lower panel}: The same as top, but for the best-fit power index at each $t_D$.
}
\label{fig:efficiency}
\end{figure}

 \subsection{Investigating the spectral variations}
We confirmed the supersoft X-ray spectra of \src which are evident in the {\it XMM} and 
\swift 2011 {observations} with sufficient counts.
These spectra can be roughly described with a thermal disk dominant emission of blackbody temperature
$\sim50$ eV plus a weak powerlaw component. 
It is important to note that the source X-ray spectrum in the {\it XMM}2014 
observation {was} getting softer when compared with the {\it XMM}2010 observation (Figure 1). 
Using the best-fitting results with the {\tt bbody+powerlaw} model (Table 2), we find that not only 
the blackbody temperature decreases in the {\it XMM}2014 data, the power-law index used to describe 
the hard X-ray emission appears to be steeper, albeit with {large errors} for the latter. 
Similar spectral softening is observed by comparing the \swift 2011 with 2013 data as well. 
The large uncertainty in temperature makes it difficult to conclude whether
the \swift 2013 spectrum {is softer than that in the {\it XMM}2014 observation, 
where the source is a factor of $\sim$2 higher in flux}.
Using the best-fitting model, we constructed two-parameter, joint confidence contours of 
the blackbody temperature versus the power-law intensity (at 1 keV) for the \swift 2011, 
{\it XMM}2010 and {\it XMM}2014 data, which are shown in Figure 4 (a). 
It can be seen that the 99\% confidence contours for the {\it XMM}2010 and {\it XMM}2014 data 
are mutually exclusive, indicating variability of the blackbody component. 
In the \swift 2011 and {\it XMM}2010 data the power-law intensity is marginal, and 
at 99\% confidence could be zero, but values of up to $4\times10^{-6}$ photons cm$^{-2}$ keV$^{-1}$ s$^{-1}$ are not ruled out. 
Figure 4 (b) shows the time evolution of hard X-ray flux in the 2-10 keV. 
Although it represents only a minor fraction of the X-ray luminosity ($<5$\%) 
and the errors are large, 
we found tentative evidence that the hard X-ray flux is declining. 

\begin{figure}[]
\centering
\includegraphics[scale=0.55,angle=0]{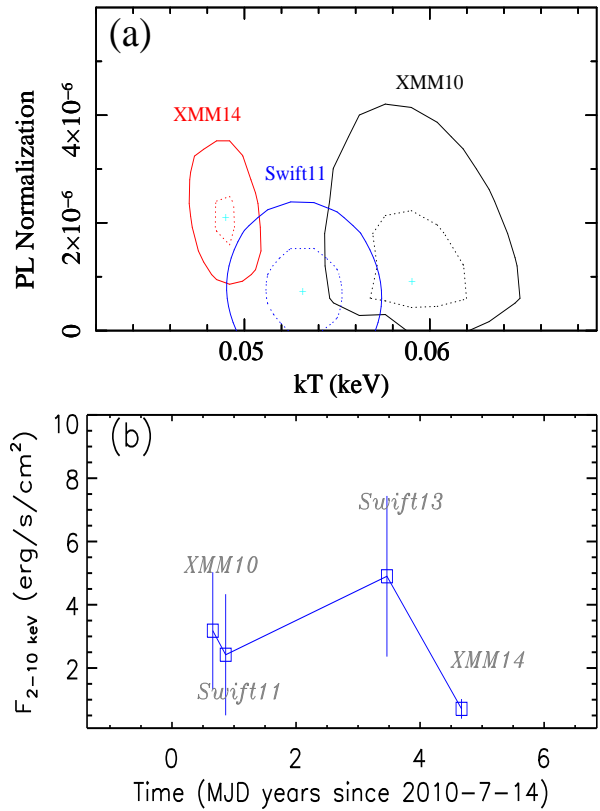}
 \caption{
 {\it Upper panel:} 
 Joint 68\% (dotted) and 99\% (solid) confidence contours of the blackbody temperature 
 vs. normalized flux for the hard power-law component. The latter is defined as flux at 1 keV 
 in units of photons cm$^{-2}$ keV$^{-1}$ s$^{-1}$. 
 Only observations that have good 
 statistics in the spectrum ({\it XMM}2010, {\it Swift}2011 and {\it XMM}2014) are shown. 
 {\it Lower panel:} The 2--10 keV flux light curve, which was estimated by fitting 
  a simple model of blackbody plus a weak powerlaw component. See Section 3.3 for details. 
 }
 \end{figure}

\section{Discussion and Conclusion}

It was established previously that the optical spectrum of \src is that of 
type II AGN (M13). 
This may suggest that the X-ray outburst could be attributed to the AGN activity. 
However, variability of more than a factor of 240 is extremely rare among AGNs. 
In particular, few AGN is known to show X-ray spectra as soft as \src at energies 
below $\sim$1.5 keV (Terashima et al. 2012; Lin et al. 2013; Sun13). 
M13 interpreted the large X-ray variability as transition from a  
radiatively inefficient state to a disk dominant thermal state, which is commonly 
observed in the XRB systems when the accretion rate is high. 
However, such a supersoft state would last more than $10^3$ years in AGNs if the analogy with 
XRBs applied. This is at odds with our finding of the slow decay in the source flux 
within $\sim$ decade. 
The radiation-pressure instability in the AGN accretion disk seems also unable to 
explain the X-ray burst in \srcs. 
{The model predicts a slow-rise and fast-decay light curve, i.e., 
the duration during the flux rising phase is much longer than that in the decay phase 
(e.g., Czerny et al. 2009). 
In addition, the outburst due to the radiation-pressure instability is expected to last 
$\sim10^2-10^3$ years for a black hole of mass $M=10^6$\msun.
All these characteristics seem to contradict to what is observed for the source.}
The outburst may be alternatively explained by a thermal-viscous instability in the accretion disk, 
as proposed for the low-luminosity AGN NGC 3599 (Saxton et al. 2015) and IC 3599 (Grupe et al. 2015). 
In this model the flares will repeat on the viscous timescale of few decades.  
Future long-term monitoring observations will enable to test this mechanism.  

Although we cannot completely rule out that the large amplitude variability is 
caused by a highly variable AGN in \srcs, it could be most likely associated with {a TDE}, 
in light of the slow flux decay, supersoft X-ray spectrum and its changes with flux. 
Theoretical works and simulations have suggested that the expected TDE rate in {AGNs} 
is much higher than the quiescent galaxy (Karas \&  Subr 2007; Kennedy et al. 2016). 
However, only a few flares/outbursts in AGNs have been considered as TDEs. 
In fact, the likely TDE candidate ASASSN14li has been detected in radio before the 
flare, which suggests the presence of pre-existing AGN activity. 
More recently, the transient PS16dtm was interpreted as a TDE in the highly accreting 
Narrow Line Seyfert 1 Galaxy (Blanchard et al. 2017).
One major difference between \src and other TDE candidates is its slow evolution in flux. 
The duration of most TDEs observed is short, whose flux decreases by an order of magnitude 
within months to {a few} years (Komossa 2015). 
On the contrary, the X-ray flux for \src dropped by only a factor of $\sim$2--4 in $\sim$8 {years} 
(2010-2017) after it was detected as an X-ray outburst. 
{The evolution at a relatively later time for \src may explain the difference.}
{We attempted to model the light curve with a $(t-t_D)^\alpha$ decline law, and found acceptable 
fits for a broad range of decay index between $\sim$-0.5 and -2.0 (Figure 3). 
The value of the decay index cannot be tightly constrained with current data, as it depends strongly 
on the TDE time, $t_D$, which is poorly known.  
A decay index of $-5/3$ would yield a $t_D=2004.98$, which is $\sim$5.6 years before the 
peak of the luminosity seen by {\it XMM}. 
Later occurring time for TDE, shallower decay index is expected.


Regardless of the actual TDE time, the slow decay in the X-ray emission and spectral variability}
make \src somewhat similar to the long-lived TDE candidate 
3XMM J150052.0+015452 (J1500+0154), which showed little decay of the X-ray flux over few tens years since outburst 
(Lin et al. 2017b). 
While tentative spectral softening has been observed in \src when the X-ray flux is lower, 
J1500+0154 displayed more clear evidence of dramatic spectral softening from quasi-soft 
($kT\sim$0.3 keV) to super-soft ($kT\sim$0.13 keV) state (Lin et al. 2017b). 
Yet no sign of persistent nuclear activity is seen in the optical emission
lines of the host galaxy for J1500+0154. 
Meanwhile, the large decay of X-ray flux has also been reported in the type II AGN 
2XMM J1231+1106 (Lin et al. 2017a), which has similar supersoft X-ray spectrum as \srcs, 
indicating the long-lasting TDE maybe relevant to supersoft AGNs. 


It is important to note that weak hard X-ray emission was detected in \srcs,  
which is most evident in the long \xmm observation performed in 2014. 
In addition, we found tentative evidence of the hard X-ray flux decreasing with 
time, the nature of which is unknown. 
Such a hard X-ray component is rarely seen in TDEs (e.g., Komossa 2015; Saxton et al. 2017), but could be 
associated with the pre-existing AGN activity that has been inferred form the 
optical spectrum for the source. 
If related to the coronal X-ray emission commonly seen in AGNs, the strength of the 
hard X-ray emission before the flare  can be roughly estimated from the [{\sc O~iii}]5007 luminosity (e.g. Lamastra et al. 2009). 
\src has a [{\sc O~iii}]5007 luminosity of $1.1\times10^{40}$\erg from the 6dF spectrum (Saxton et al. 2011), 
which implies a 2--10 keV luminosity of $1.1\times10^{41}$\erg (the dispersion is 0.63 dex). 
This is a factor of five higher than that highest flux measured with \xmm 2010 data, 
indicating a difference in the hard X-ray flux before and after the flare. 

{ 
It is possible that the accretion of stellar debris from TDE interacts with the pre-existing corona
from which the hard X-rays originate, causing a change in the structure and scattering optical depth 
in the X-ray emitting region. 
This may explain both the spectral and flux variations of the hard powerlaw component in this
object. 
Similar scenario has been proposed to explain the X-ray flux drop in the transient PS16dtm, as being due 
to the obscuration of the pre-existing AGN corona by the stellar debris from TDE (Blanchard et al. 2017). 
However, since the hard X-ray spectra are still poor for most of observations due to few counts detected, 
it is not yet completely clear whether the hard X-ray emission in GSN 069
is produced in pre-existing AGN, a new-formed corona associated with TDE, or jet. 
We have obtained observing time from JVLA which will provide new insight into the
origin of the hard X-ray emission through the detection of jet-like radio emission or not. }







\begin{acknowledgements}
We thank the Swift Acting PI, Brad Cenko, for approving
our ToO request to observe \srcs.
This research made use of the HEASARC online data archive
services, supported by NASA/GSFC. 
This work is supported by Chinese NSF through grant 11233002, 11573001, 116203021, u1731104, and   
National Basic Research Program 2015CB857005.
X.S. acknowledges 
the support from Anhui Provincial NSF (1608085QA06) and Young Wanjiang Scholar program. 
\end{acknowledgements}




\end{document}